\documentclass[12pt]{article}
\usepackage{graphicx}
\usepackage{subfigure}
\usepackage{amsfonts}
\usepackage{amssymb,amsmath}
\usepackage{hyperref}
\usepackage{comment}
\usepackage{multicol}
\usepackage[usenames,dvipsnames]{xcolor}
\definecolor{rosso}{cmyk}{0,1,1,0.3}
\definecolor{verde}{cmyk}{0.8,0,0.6,0.25}
\definecolor{bluc}{cmyk}{1,0.4,0,0.1}
\definecolor{blucc}{cmyk}{0.8,0.3,0,0}

\def\mdm{m_{\rm DM}}

\newcommand{\lrfp}[3]{ \left(\frac{#1}{#2}\right)^{#3} }

\begin{document}
\begin{titlepage}
\begin{flushright}
UT-14-7\\
IPMU 14-0045\\
UG-FT308/14\\
CAFPE178/14
\end{flushright}

\vskip 3cm

\begin{center}

{\Large\bf Isospin--Violating Dark Matter\\ \vspace{.3em} with Colored Mediators}

\vskip .5in

{
Koichi Hamaguchi$^{(a,b)}$,
Seng Pei Liew$^{(a)}$, \\
Takeo Moroi$^{(a,b)}$
and
Yasuhiro Yamamoto$^{(c)}$
}

\vskip .3in

{\em
$^a$Department of Physics, University of Tokyo, Bunkyo-ku, Tokyo 113-0033, Japan \vspace{0.2cm}\\
$^b$Kavli Institute for the Physics and Mathematics of the Universe,
University of Tokyo, Kashiwa 277-8583, Japan \\
$^c$Departamento de F\'{\i}sica Te\'orica y del Cosmos, Universidad de Granada, Granada~E-18071,~Spain
}

\begin{abstract}
  In light of positive signals reported by the CDMS--II Si experiment
  and the recent results of the LUX and SuperCDMS experiments, we study
  isospin--violating dark matter scenarios assuming that the
  interaction of the dark matter is mediated by colored particles.  We
  investigate the phenomenology of the model, including collider
  searches, flavor and CP phenomenology.  A minimal possible scenario
  includes scalar dark matter and new vector-like colored fermions
  with masses of O(1) TeV as mediators.  Such a scenario may be probed
  at the 14 TeV LHC, while flavor and CP constraints are stringent and
  severe tuning in the couplings is unavoidable.  We also found that, as an explanation of the CDMS--II Si signal, isospin--violating fermionic dark matter models with colored scalar mediators are
  disfavored by the LHC constraints.
\end{abstract}

\end{center}
\end{titlepage}


\section{Introduction}
Dark matter (DM), which is expected to be responsible for about $27\
\%$ of the mass density of the present universe \cite{Ade:2013zuv}, is
still a great mystery to the field of particle physics.  Although
various cosmological observations have confirmed the existence of DM,
its particle-physics properties, such as the mass, strength of its
interactions with Standard--Model (SM) particles, and so on, remain fully
unknown.  Various experiments have been performed to detect direct and
indirect signals of DM~\cite{Bertone:2004pz}.

In recent years, several direct detection experiments
(DAMA/LIBRA~\cite{Bernabei:2010mq},
CoGeNT~\cite{Aalseth:2010vx,Aalseth:2011wp,Aalseth:2014eft},
CRESST~\cite{Angloher:2011uu} and the CDMS-II Si
experiment~\cite{Agnese:2013rvf}) have found signals
that may suggest the existence of
light DM with mass around 10 GeV.  On the other hand,
experiments such as~XENON~\cite{Angle:2011th,Aprile:2012nq},
LUX~\cite{Akerib:2013tjd}, SIMPLE~\cite{Felizardo:2011uw},
CDMS~\cite{Akerib:2010pv,Ahmed:2010wy}
and
SuperCDMS~\cite{Agnese:2013jaa, Agnese:2014aze} 
have not found any excess of events
that can be interpreted as signals from DM.  In particular,
the LUX experiment has probed the relevant region of parameters at the
highest level of sensitivity and excluded most regions favored by the
possible signals of light DM.

It has been shown that it is difficult to accommodate positive signals of direct detection experiments and bounds from Xenon-based experiments by considering astrophysical alternatives (e.g., modified halo models) or varying assumptions about the Xenon scintillation efficiencies~\cite{Gresham:2013mua,DelNobile:2013gba,Fox:2013pia}.
A scenario that still remains viable in reconciling some of these results is the
isospin--violating DM~\cite{Kurylov:2003ra,Giuliani:2005my,Chang:2010yk,Kang:2010mh,Feng:2011vu,Frandsen:2013cna,Feng:2013vod}.
As different types of nuclei are used in different direct detection experiments, isospin--dependent interactions may happen to interfere destructively for a certain type of nuclei, and thus suppress the DM--nucleus scattering cross section.
As LUX experiment~\cite{Akerib:2013tjd} currently imposes the most stringent bound on DM, one necessarily considers DM that has negligible interaction with the Xenon nucleus.
Recent studies after the LUX result~\cite{Gresham:2013mua,DelNobile:2013gba,Cirigliano:2013zta,Fox:2013pia}
have shown that the isospin--violating DM is still 
compatible with one of positive signals, 
those of the CDMS-II Si experiment.

More recently, SuperCDMS Collaboration reported their first result for the WIMP search
using their background rejection capabilities~\cite{Agnese:2014aze}.
As we shall see, the isospin--violating DM scenario
is severely constrained also by SuperCDMS, but there is still a viable region of parameter
space.
    
In this paper, we study a minimal extension of the SM with
isospin--violating DM, assuming that the isospin--violating
interaction of the DM is mediated by colored particles.\footnote{For
  recent studies on DM models with colored mediators, see, e.g.,
  Refs.~\cite{Chang:2013oia,An:2013xka,Bai:2013iqa,DiFranzo:2013vra,Papucci:2014iwa}.}
We investigate the phenomenology of the model, including collider
searches as well as flavor and CP physics, paying particular attention
to the parameter region which is consistent with CDMS--Si, LUX and SuperCDMS
results. We show that a minimal viable model includes scalar DM and
new colored vector-like fermions with masses of $O(1)$ TeV as
mediators.  The colored vector--like fermions can be tested at the 14
TeV LHC.  On the other hand, the flavor and CP constraints severely
restrict the parameters of the model.  We also show that fermionic DM
models with colored scalar mediators are disfavored by the LHC
constraints.

The remaining of the paper is organized as follows. In Section~\ref{s:eff}, we study effective operators involving SM and DM fields that reproduce the direct detection experimental results. 
We then study bounds on these operators from collider search and indirect detection. 
In Section~\ref{s:main}, we introduce a simple model  involving only DM and colored mediators as new particles, that can reproduce the effective operators 
studied in Section~\ref{s:eff}.
We study the current bound (8 TeV LHC) on the colored mediators and their prospects of discovery for 14 TeV LHC. In Section~\ref{SecFlavor}, we examine flavor and CP constraints on this model. Section~\ref{sec:conclusion} is devoted to conclusions. 
In Appendix~\ref{s:maj}, we briefly discuss models of isospin--violating fermionic DM mediated by colored scalars.
          

\section{Effective Operators of Isospin--Violating Dark Matter}
\label{s:eff}

\subsection{Direct detection}

We study the case in which the DM interaction is dominated by spin-independent interaction.\footnote{It is difficult to interpret the CDMS-Si signal as spin-dependent scattering of isospin-violating DM~\cite{Buckley:2013gjo}.
}~In the non-relativistic limit, the elastic scattering cross section of DM with a nucleus composed of $Z$ protons and $(A-Z)$ neutrons can be represented as
\begin{align}
\sigma_A &\simeq  \frac{\mu_A^2}{4 \pi \mdm^2}\left[ f_p Z + f_n (A-Z) \right]^2,
\label{eq:sigmaA}
\end{align}
where $\mu_A=m_A \mdm/(m_A + \mdm)$ is the reduced mass, $m_A$ is
the mass of the nucleus and $\mdm$ is the mass of the DM.  $f_n$ and
$f_p$ parametrize the coupling between DM and neutron and proton,
respectively.  Their explicit forms in terms of Lagrangian parameters
are shown in the following subsections.  An isospin-conserving
interaction corresponds to $f_n=f_p$.  If the isospin is violated and
the ratio of the couplings satisfy a relation $f_n/f_p\simeq
-Z/(A-Z)$, the cross section $\sigma_A$ is suppressed.  In particular,
the DM--Xenon interaction is suppressed for $f_n/f_p\simeq -0.7$.

Given the very severe bound from LUX experiment, it is important to include the effects of multiple isotopes~\cite{Feng:2011vu}, which leads to
\begin{align}
\sigma_A &= \frac{1}{4 \pi \mdm^2}\sum_i\eta_i \mu_{A_i}^2\left[ f_p Z + f_n (A_i-Z) \right]^2,
\end{align}
where $\eta_i$ is the natural abundance of the $i$-th isotope.
Results of direct detection experiments are often quoted in terms of ``normalized-to-nucleon cross section,"
which is given by
\begin{align}
\sigma_N^{(Z)} &= \frac{\mu_p^2}{\sum_i \eta_i \mu_{A_i}^2 A_i^2} \sigma_{A}
=\frac{\sum_i \eta_i \mu_{A_i}^2 \left[ Z + (f_n/f_p) (A_i - Z)\right]^2}
{\sum_i \eta_i \mu_{A_i}^2 A_i^2}
\sigma_p\,.
\label{eq:sigma_p}
\end{align}
In the isospin conserving case, $f_n = f_p$, this is equal to DM--proton cross section $\sigma_p$.
The DM--Xenon scattering cross section is minimized for $f_n/f_p\simeq -0.7$.

\begin{figure}[tbp]
 \begin{center}
 \includegraphics[width=10cm]{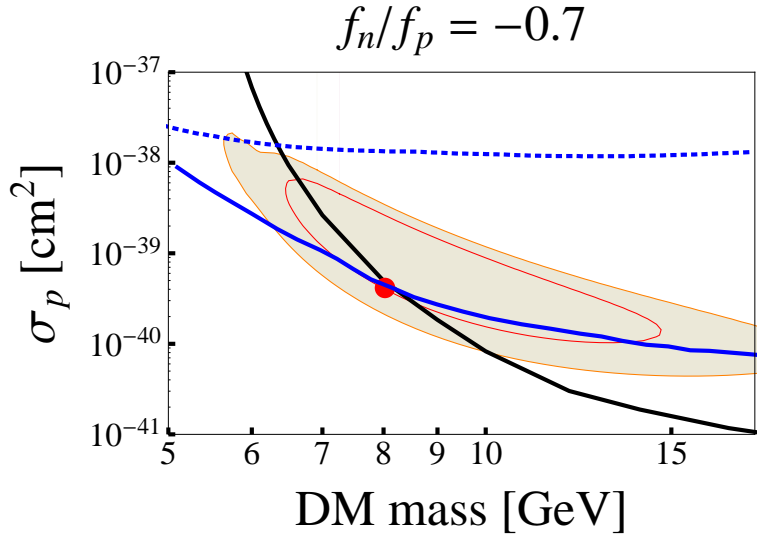}
 \end{center}
 \caption{Favored and excluded regions in isospin-violating DM with
   $f_n/f_p=-0.7$.  Shaded regions show 68\% and 90\% confidence level
   contours for a possible signal from the CDMS-Si
   result~\cite{Agnese:2013rvf}. Black solid, blue dashed, and blue solid lines
   represent the exclusion contours from LUX~\cite{Akerib:2013tjd},
   CDMSlite~\cite{Agnese:2013jaa}, and the recent SuperCDMS
   \cite{Agnese:2014aze} experiments, respectively.
   The red point represent the benchmark point used in our analysis.  
   }
\label{fig:CDMSvsLUX}
\end{figure}

In Fig.~\ref{fig:CDMSvsLUX}, we show the parameter regions in $(\mdm,
\sigma_p)$ plane for $f_n/f_p=-0.7$, which are favored by CDMS-Si, and
excluded by LUX and SuperCDMS.\footnote{Among many direct detection
  experimental results, we show in Fig.~\ref{fig:CDMSvsLUX} only
  positive signals from CDMS--Si and bounds from LUX and
  SuperCDMS.  
  This is because LUX and SuperCDMS give the most stringent
  constraints on isospin--violating DM with $f_n/f_p=-0.7$, and only
  CDMS--Si has significant region of parameters that is not excluded
  by these bounds. 
}  In the following analysis, we consider the following 
representative point of isospin--violating DM:
\begin{align}
\mdm=8~\text{GeV},
\;
\sigma_p=4\times 10^{-40}\text{cm}^2,
\; 
f_n/f_p = -0.7.
\label{eq:representative_point}
\end{align}
One can see that this point is marginally allowed by LUX and
SuperCDMS, and is favored by CDMS-Si.  

\subsection{Effective interactions between quarks and dark matter}
\begin{table}
\begin{center}
\begin{tabular}{|l|l|l|}
\hline
DM & operator ${\cal O}$ & $f_N$
\\ \hline
real  & 
${\cal O}^{\text{(R)}} \equiv
\displaystyle \sum_{q=u,d} C_q^{\text{(R)}} \frac{1}{2}\phi^2 \cdot \bar{q}q$
& $\displaystyle   \sum_{q=u,d} B_q^{(N)} C_q^{(\text{R})}$
\\
complex  (S) &
${\cal O}^{\text{(Cs)}} \equiv
\displaystyle \sum_{q=u,d} C_q^{\text{(Cs)}} \phi^* \phi \cdot \bar{q}q$
& $\displaystyle   \sum_{q=u,d} B_q^{(N)} C_q^{(\text{Cs})}$
\\
complex  (V) &
${\cal O}^{\text{(Cv)}} \equiv
\displaystyle \sum_{q=u,d} C_q^{\text{(Cv)}} i 
(\phi^* \partial_\mu \phi - \phi \partial_\mu \phi^*) 
\bar{q}\gamma^\mu q$
& $\displaystyle  2 m_{\text{DM}}\times
\begin{cases}
2C_u^{\text{(Cv)}} + C_d^{\text{(Cv)}}\;\;(f_p)
\\
C_u^{\text{(Cv)}} + 2C_d^{\text{(Cv)}}\;\;(f_n)
\end{cases}$
\\ \hline
Majorana  &
${\cal O}^{\text{(M)}} \equiv
\displaystyle \sum_{q=u,d} C_q^{\text{(M)}} \frac{1}{2}\bar{\chi}\chi \cdot \bar{q}q$
& $\displaystyle 2 m_{\text{DM}} \sum_{q=u,d} B_q^{(N)} C_q^{(\text{M})}$
\\
Dirac  (S) &
${\cal O}^{\text{(Ds)}} \equiv
\displaystyle \sum_{q=u,d} C_q^{\text{(Ds)}} \bar{\chi}\chi \cdot \bar{q}q$
& $\displaystyle  2 m_{\text{DM}} \sum_{q=u,d} B_q^{(N)} C_q^{(\text{Ds})}$
\\
Dirac  (V) &
${\cal O}^{\text{(Dv)}} \equiv
\displaystyle \sum_{q=u,d} C_q^{\text{(Dv)}} \bar{\chi}\gamma_\mu \chi \cdot \bar{q}\gamma^\mu q$
& $\displaystyle  2 m_{\text{DM}}\times
\begin{cases}
2C_u^{\text{(Dv)}} + C_d^{\text{(Dv)}}\;\;(f_p)
\\
C_u^{\text{(Dv)}} + 2C_d^{\text{(Dv)}}\;\;(f_n)
\end{cases}$
\\ \hline
\end{tabular}
\end{center}
\caption{Effective operators of quark--DM interactions.}
\label{tab:operators}
\end{table}

The particle DM can be a real or complex scalar field~$\phi$
(and~$\phi^*$ if complex). It can also be a Majorana or a Dirac
fermion~$\chi$.  Assuming that the scattering with nucleon is
dominated by spin-independent interaction, there exist only six
effective operators at the quark level as listed in
Table~\ref{tab:operators}.
\footnote
{In the present scenario, the energy scales relevant for
    collider physics, dark matter detection, and CP / flavor physics
    are different.  In our calculation, however, renormalization group
    effects on the Wilson coefficients are neglected.}
  In this section, we consider only the
couplings of DM to up- and down-type quarks, since they give the
dominant isospin--violating effects.  In the table, we also express
the parameters $f_n$ and $f_p$ in terms of the effective couplings
$C_q$, where $B^{(N)}_q = \langle N | \bar{q}q | N\rangle = m_N
f_{T_q^{(N)}}/m_q$ ($N=p, n$) are neutron and proton scalar matrix
elements.  In our numerical analysis, we use the following values:
$B_d^{(p)}/B_u^{(p)} = B_u^{(n)}/B_d^{(n)} = 0.80$~\cite{Ellis:2008hf}
and $B_u^{(p)}+B_d^{(p)} = B_u^{(n)}+B_d^{(n)} = 2\sigma_{\pi N}/(m_u
+ m_d)$ with $\pi$-nucleon sigma term $\sigma_{\pi N}\simeq 64$
MeV~\cite{Ellis:2008hf} and light quark mass $(m_u + m_d)/2\simeq 3.5$
MeV~\cite{Beringer:1900zz}.

In order to reproduce the DM--nucleon cross section of the representative
point in Eq.~(\ref{eq:representative_point}), the effective couplings
$C_q$ in Table~\ref{tab:operators} for each scenario are determined as:
\begin{align}
C_{u}^{\text{(R,Cs)}} 
&\simeq -1.04\times C_{d}^{\text{(R,Cs)}} \simeq
(68~\text{TeV})^{-1},
\label{eq:Cu1}
\\
C_{u}^{\text{(M,Ds)}} 
&\simeq -1.04\times C_{d}^{\text{(M,Ds)}} \simeq 
(1050~\text{GeV})^{-2},
\label{eq:Cu2}
\\
C_{u}^{\text{(Cv,Dv)}} 
&\simeq -1.13\times C_{d}^{\text{(Cv,Dv)}} 
\simeq 
(720~\text{GeV})^{-2}.
\label{eq:Cu3}
\end{align}


\subsection{LHC bounds on the effective operators}
\label{sec:LHC:eff}


The DM-SM effective operator approach applied to collider physics has
been useful in complementing direct and indirect probes of DM~\cite{Birkedal:2004xn, Beltran:2010ww, Goodman:2010yf,Goodman:2010ku,Bai:2010hh,
  Fox:2011fx, Fox:2011pm, Bai:2012xg}. 
When the DM production at colliders is accompanied by 
a jet from initial state radiation, the 
signature will be a jet (mono--jet) with missing
transverse energy (MET).  The ATLAS and CMS collaborations have also
performed searches on mono--photon plus MET, mono--lepton and
mono--$W$ or --$Z$ plus MET.  These searches currently
provide the most stringent collider bounds on DM~\cite{Chatrchyan:2012pa, cms-dm, Chatrchyan:2012tea, cms-dml,
  ATLAS:2012ky, atlas-dm2, Aad:2013oja}.  To see how severely the
isospin--violating DM model is constrained by the LHC data,
we have calculated the cross section for these processes.  Here, we
assume that one of the operators listed in Table~\ref{tab:operators}
dominates the signal process.

For the mono--$W$ and --$Z$ events, our analysis is based on the
ATLAS study given in Ref.~\cite{Aad:2013oja},
which utilizes the hadronic decay modes of $W$ and $Z$ boson.  We have
generated the signal events using~{\sc MadGraph 5}~\cite{Alwall:2011uj},
assuming the existence of one of
the operators given in Table~\ref{tab:operators}.  We apply, in
accord with~\cite{Aad:2013oja}, the following cuts at the parton level:
\begin{itemize}
\item 
$p_T^{W, Z} > 250$ GeV,
where $p_T^{W, Z}$ is the transverse momentum of $W$ or $Z$,
\item
$|\eta|^{W, Z} < 1.2$,
where $\eta^{W, Z}$ is the pseudo-rapidity,
\item
$\sqrt{y} > 1.2$, where $\sqrt{y}={\rm min}(p_{T1},p_{T2})\Delta
  R/m_{\rm jet}$, with $p_{Ti} (i=1\ {\rm or}\ 2)$ being the
  transverse momentum of jet from the decay of $W$ or $Z$, $\Delta R$
  the distance between jets, and $m_{\rm jet}$ the calculated mass of
  the jet.
\end{itemize}
The fiducial efficiency (63 \%) has been taken into account as well.
Upper bounds on the dimensionful couplings of the effective operators
in Table\ \ref{tab:operators} are obtained based on the observed
upper limits on the cross section at 95 \% CL in Ref.~\cite{ATLAS:2012ky}.

We have also calculated the cross section for the mono--jet events.
(For the mono--jet bounds on isospin--violating DM, see also
\cite{Kumar:2011dr,Feng:2013vod}.)  To make a comparison with the
ATLAS mono--jet search at 7 TeV~\cite{ATLAS:2012ky},\footnote
{Results at 8 TeV~\cite{atlas-dm2} do not have significant
  improvements compared to the limits obtained in
  \cite{ATLAS:2012ky}.}
we calculate the parton-level cross section with the
following cuts on  the mono--jet momentum:
\begin{itemize}
\item $p_T > 80$ GeV,
\item $|\eta| < 2.0$.
\end{itemize}
The parton-level cross section is multiplied by the signal acceptance.
(Here, we also include the efficiency of the detector, which is taken
to be 83 \% \cite{ATLAS:2012ky}.)  In \cite{ATLAS:2012ky}, the signal
acceptance for the cases with ${\cal O}^{\text{(R)}}$, ${\cal
  O}^{\text{(Cs)}}$, ${\cal O}^{\text{(M)}}$ and ${\cal
  O}^{\text{(Ds)}}$ are not presented.  For these cases, we use the
acceptance for D5 model given in \cite{ATLAS:2012ky}.  (Notice that
the scalar interactions considered in \cite{ATLAS:2012ky}, i.e., D1
model, are proportional to the quark masses and the effect of
$c$-quark is important.  Thus, we do not use the acceptance of the D1
model in our analysis.)  We found that, among several signal
regions~\cite{ATLAS:2012ky}, the one corresponding to $p_T > 350$ GeV
(SR3) gives the most stringent bounds to the present model.  Comparing
our estimations of the cross sections with the observed 95 \% CL limit
on the ``visible cross section'' given in \cite{ATLAS:2012ky}, we
derive upper bounds on the coefficients of the effective operators
listed in Table~\ref{tab:operators}.

The bounds are given in Table~\ref{tab:mono}.
Here, we show the results based on the mono--$W$ and $Z$ events and
mono--jet events separately.  We can see that the mono--$W$ and --$Z$
processes impose more stringent constraints than
the mono-jet process.  One of the reasons is that in the
isospin--violating DM model, there exists the relative minus
sign between the coupling of DM to $u$- and $d$-quarks; it
results in a constructive interference between two Feynman diagrams
for the mono--$W$ production process that greatly enhances the cross
section \cite{Bai:2012xg}.

\begin{table}[t] 
\centering 
\begin{tabular}{|c|c|c|}
\hline
Effective operator &mono--$W$ or $Z$&mono--jet
\\
\hline
$C^{\text{(R)}}_u$&(630 GeV)$^{-1}$&(400 GeV)$^{-1}$
\\
\hline
$C^{\text{(Cs)}}_u$&(890 GeV)$^{-1}$&(570 GeV)$^{-1}$
\\
\hline
$C^{\text{(M)}}_{u}$&(820 GeV)$^{-2}$&(470 GeV)$^{-2}$
\\
\hline
$C^{\text{(Ds)}}_u$&(970 GeV)$^{-2}$&(560 GeV)$^{-2}$
\\
\hline
$C^{\text{(Cv)}}_u$&(760 GeV)$^{-2}$&(430 GeV)$^{-2}$
\\
\hline
$C^{\text{(Dv)}}_u$&(1100 GeV)$^{-2}$&(610 GeV)$^{-2}$
\\
\hline
\end{tabular} 
\caption{Upper bounds on the coefficients of 
the effective operators
obtained from mono--$W$ or --$Z$
  and   mono--jet searches.
}
\label{tab:mono} 
\end{table} 

Comparing Eqs.~(\ref{eq:Cu1})--(\ref{eq:Cu3}) with
Table~\ref{tab:mono}, the CDMS-Si point with the vector--type
effective operators ${\cal O}^{\text{(Cv)}}$ and ${\cal
  O}^{\text{(Dv)}}$ are disfavored.  
On the other hand, scalar--type interactions, 
${\cal O}^{\text{(R)}}$, ${\cal O}^{\text{(Cs)}}$, ${\cal O}^{\text{(M)}}$ 
and ${\cal O}^{\text{(Ds)}}$, are
still viable.  In the next section we introduce a simple model which
can reproduce the effective operators ${\cal O}^{\text{(R)}}$ and ${\cal O}^{\text{(Cs)}}$
at low energy.
(For fermionic DM with effective operators ${\cal O}^{\text{(M)}}$ and ${\cal O}^{\text{(Ds)}}$, see Appendix~\ref{s:maj}.)

Before closing this section, let us comment on the validity of the
effective field theory (EFT) approach.  The effective operators at low
energy are generated by a UV theory, typically by exchanges of heavy
mediators.  At the LHC, the energy scale of the process can be
comparable to or larger than the scale of the UV theory.  In such a
case, the bound obtained by using EFT may not be valid~\cite{Friedland:2011za,Shoemaker:2011vi,Busoni:2013lha,Chang:2013oia,An:2013xka,Bai:2013iqa,DiFranzo:2013vra,Buchmueller:2013dya,Papucci:2014iwa}.  However, when
the effective operators are induced by exchanges of heavy colored mediators, 
the bound obtained by EFT is
typically weaker than the bound obtained by concrete UV models, i.e.,
EFT gives conservative
bounds~\cite{Papucci:2014iwa}. Thus, we consider the
constraints obtained in this subsection as conservative ones, and
discuss UV models for the operators which are not disfavored at the
level of EFT.

\subsection{Thermal abundance and indirect search}

Another important check point is the relic abundance.  Although we
have assumed the correct DM abundance, the thermal relic
density in our model is larger than the present DM
density. The thermal relic density is determined by the
thermally-averaged pair annihilation cross section $\langle
\sigma_{\rm ann} v_{\rm rel} \rangle$ as
\begin{align}
  \Omega_{\rm thermal} \simeq 0.2\times 
  \left( 
    \frac{\langle \sigma_{\rm ann} v_{\rm rel} \rangle}{1\ {\rm pb}}
  \right)^{-1}.
  \label{Omega_th}
\end{align}

We show $\langle \sigma_{\rm ann} v_{\rm rel} \rangle$ in Table
\ref{tab:sigv} for the cases where the $s$-wave annihilation processes
dominate.  (For other cases, the annihilation is via $p$-wave
processes, with which the cross sections are much smaller.)
Substituting the cross sections in the table into Eq.\
\eqref{Omega_th}, we can see that $\Omega_{\rm thermal}$ becomes
larger than the present density parameter of DM.  Thus, we
need to consider non-thermal production of DM at $T\ll m_{\rm
  DM}$ in the present scenario.

\begin{table}
\begin{center}
\begin{tabular}{|l|l|l|}
\hline
Operator & $\langle \sigma_{\rm ann} v_{\rm rel} \rangle$
 & value
\\ \hline
${\cal O}^{\text{(R)}}$ & 
$3\sum_q |C_q^{\rm (R)}|^2 / 4\pi$  & 0.04 pb
\\ \hline
${\cal O}^{\text{(Cs)}}$ & 
$3\sum_q |C_q^{\rm (Cs)}|^2 / 8\pi$  & 0.02 pb
\\ \hline
${\cal O}^{\text{(Dv)}}$ & 
$3\sum_q |C_q^{\rm (Dv)}|^2 m_{\rm DM}^2 / 2\pi$  & 0.08 pb
\\ \hline
\end{tabular}
\end{center}
\caption{Thermally averaged total pair annihilation cross 
  sections for the cases with the operators ${\cal O}^{\text{(R)}}$, 
  ${\cal O}^{\text{(Cs)}}$, and ${\cal O}^{\text{(Dv)}}$. 
  For other cases, the cross sections are
  $p$-wave suppressed, and are much smaller.}
\label{tab:sigv}
\end{table}

We also comment on the upper bound on $\langle \sigma_{\rm ann} v_{\rm
  rel} \rangle$ from the observations of Milky-Way satellites by Fermi
Large Area Telescope (LAT)~\cite{Ackermann:2011wa,
  GeringerSameth:2011iw, Kumar:2011dr, Ackermann:2013yva}.  With the
latest analysis of the $\gamma$-ray flux from the satellites
\cite{Ackermann:2013yva}, the observed upper bound on the pair
annihilation cross section into $u\bar{u}$ or $d\bar{d}$ final state
is about 0.8 pb for $m_{\rm DM}=10\ {\rm GeV}$.  As one
can see in Table \ref{tab:sigv}, the annihilation cross section in the
present model is an order of magnitude smaller
than the Fermi-LAT bound.

\section{Colored Mediators of Isospin--Violating Dark Matter}
\label{s:main}

As we have seen in the previous section, isospin--violating DM with
scalar--type interaction can explain the possible CDMS-Si signal while
avoiding the LUX and SuperCDMS constraints as well as the LHC mono--jet and
mono--$W$/$Z$ constraints.  In this section we discuss the UV
completion of the scalar--type effective couplings.  In particular, as
mentioned in Introduction, we concentrate on the case that the
effective operators are induced by exchanges of heavy colored
particles, since they can easily accommodate isospin--violating
interactions.  For recent studies on other possibilities of
isospin--violating DM models, see, for
example,~\cite{He:2013suk,Okada:2013cba,Belanger:2013tla}.

As shown in Appendix~\ref{s:maj}, fermionic DM models require a light
colored scalar with a mass smaller than $O(500)~\text{GeV}$, and such
a model is already excluded by LHC squark search~\cite{atlas-sq}. 
 Thus, in the following discussion, we concentrate on
real and complex scalar DM.

\subsection{Model}

We introduce extra vector-like quarks $Q$, $U$, and $D$, which mediate
the coupling between scalar DM and the SM quarks. The matter content
and their quantum numbers are summarized in
Table~\ref{tab:matter_content}.  (We also list the SM
quarks to fix the notation.)  We impose a $Z_2$ symmetry to ensure the
stability of DM.
\begin{table}[t]
\begin{center}
\begin{tabular}{|c|c|c|}
\hline
particles & $SU(3)_c \times SU(2)_L \times U(1)_Y$ & $Z_2$
\\ \hline
$\phi$ & $(1,1)_0$ & $-$
\\ \hline
$Q\; (\ni Q_L, Q_R)$ & $(3,2)_{1/6}$ & $-$
\\
$U\; (\ni U_L, U_R)$ & $(3,1)_{2/3}$ & $-$
\\
$D\; (\ni D_L, D_R)$ & $(3,1)_{-1/3}$ & $-$
\\ \hline
$q_L$ &$(3,2)_{1/6}$ & $+$
\\
$u_R$ & $(3,1)_{2/3}$ & $+$
\\
$d_R$ & $(3,1)_{-1/3}$ & $+$
\\ \hline
\end{tabular}
\end{center}
\caption{Quantum numbers of DM, colored 
  mediators $Q$, $U$, and $D$ 
  (and SM quarks $q_L$, $u_R$, and $d_R$).}
\label{tab:matter_content}
\end{table}
The mass and interaction terms of the new colored fields are given by
\begin{align}
-{\cal L}_{Q,U,D}
=&
M_Q \bar{Q}Q 
+ M_U \bar{U}U
+ M_D \bar{D}D
\nonumber\\
& + 
\left(
\lambda_Q^i\; \phi\; \overline{q_L^i} P_R Q 
  +\lambda_U^i\; \phi\; \overline{u_R^i} P_L U
  +\lambda_D^i\; \phi\; \overline{d_R^i} P_L D 
\right.  
\nonumber \\
& 
+y_{U_L} H^\dagger_a \overline{Q_a} P_L U
+y_{D_L} \epsilon^{ab} H_a \overline{Q_b} P_L D
\nonumber\\
& 
\left.
+y_{U_R} H^\dagger_a \overline{Q_a} P_R U
+y_{D_R} \epsilon^{ab} H_a \overline{Q_b} P_R D
\; +\text{H.c.}\right).
\label{eq:L}
\end{align}
The index $i$ stands for the generation of SM quarks.  
Note that Yukawa couplings between
colored mediators and Higgs field is necessary
in order to induce a scalar type effective operator
between DM and SM quarks,  
${\cal L}_{\text{eff}}\sim \phi\phi\bar{q}q$.
Therefore, after the electroweak symmetry is broken,
the two up-type colored mediators mix with a mass matrix
\begin{equation}
{\cal M_U}
=\left(\begin{array}{cc}
M_U 
& 
y_{U_L}^* v
\\ 
y_{U_R} v
& 
M_Q
 \end{array}\right),
\end{equation}
where $v\simeq 174$ GeV is the Higgs vacuum expectation value.  The
two down-type mediators also mix in a similar way.

In the case of complex scalar DM, we impose a global U(1) symmetry,
where only the DM and colored mediators are charged; the Lagrangian
(\ref{eq:L}) has such a symmetry.  As we will see, this U(1) symmetry
makes phenomenology of complex scalar DM and real one different.

\begin{figure}[t]
 \begin{center}
 \includegraphics[width=7cm]{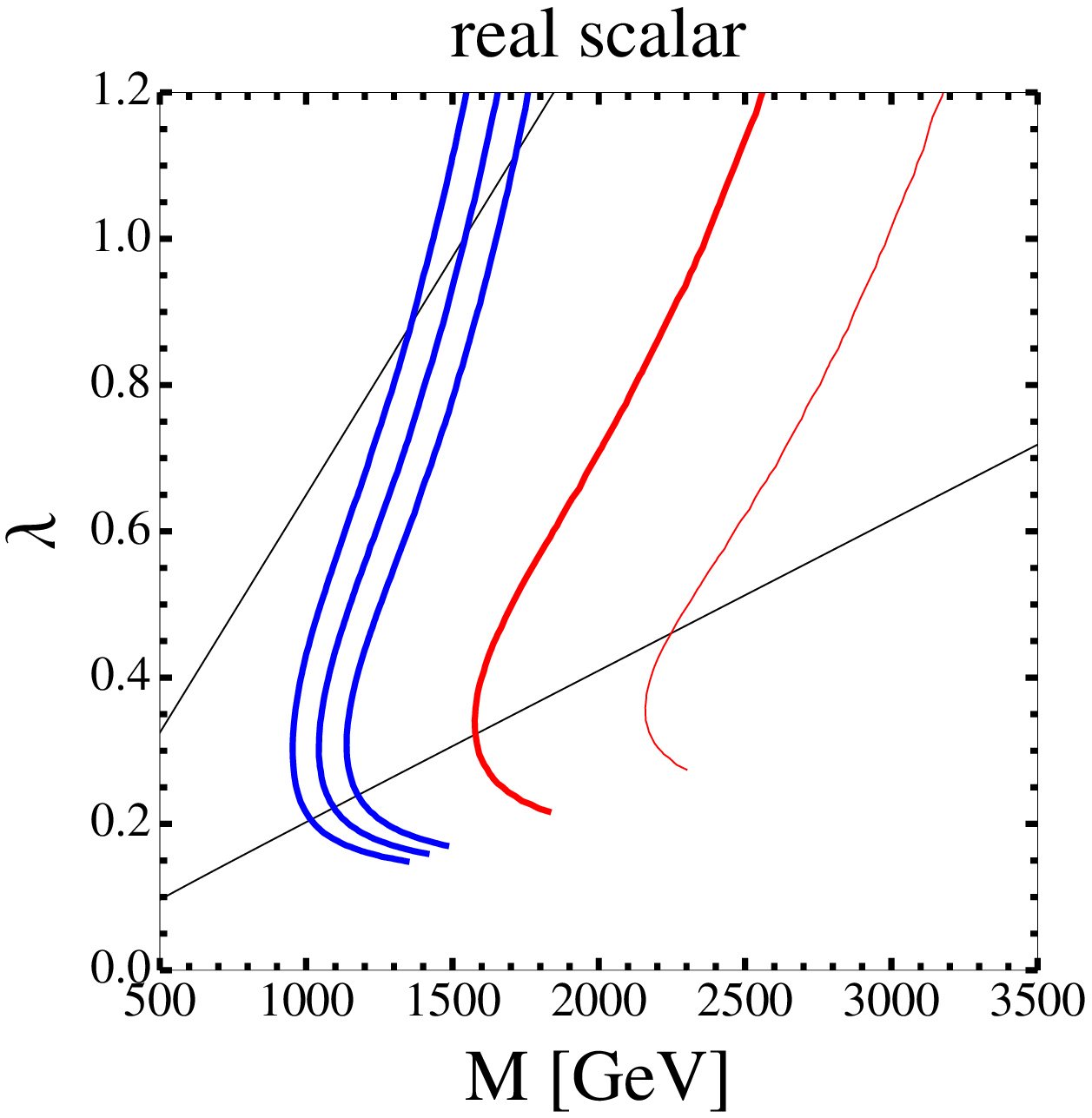}
 \includegraphics[width=7cm]{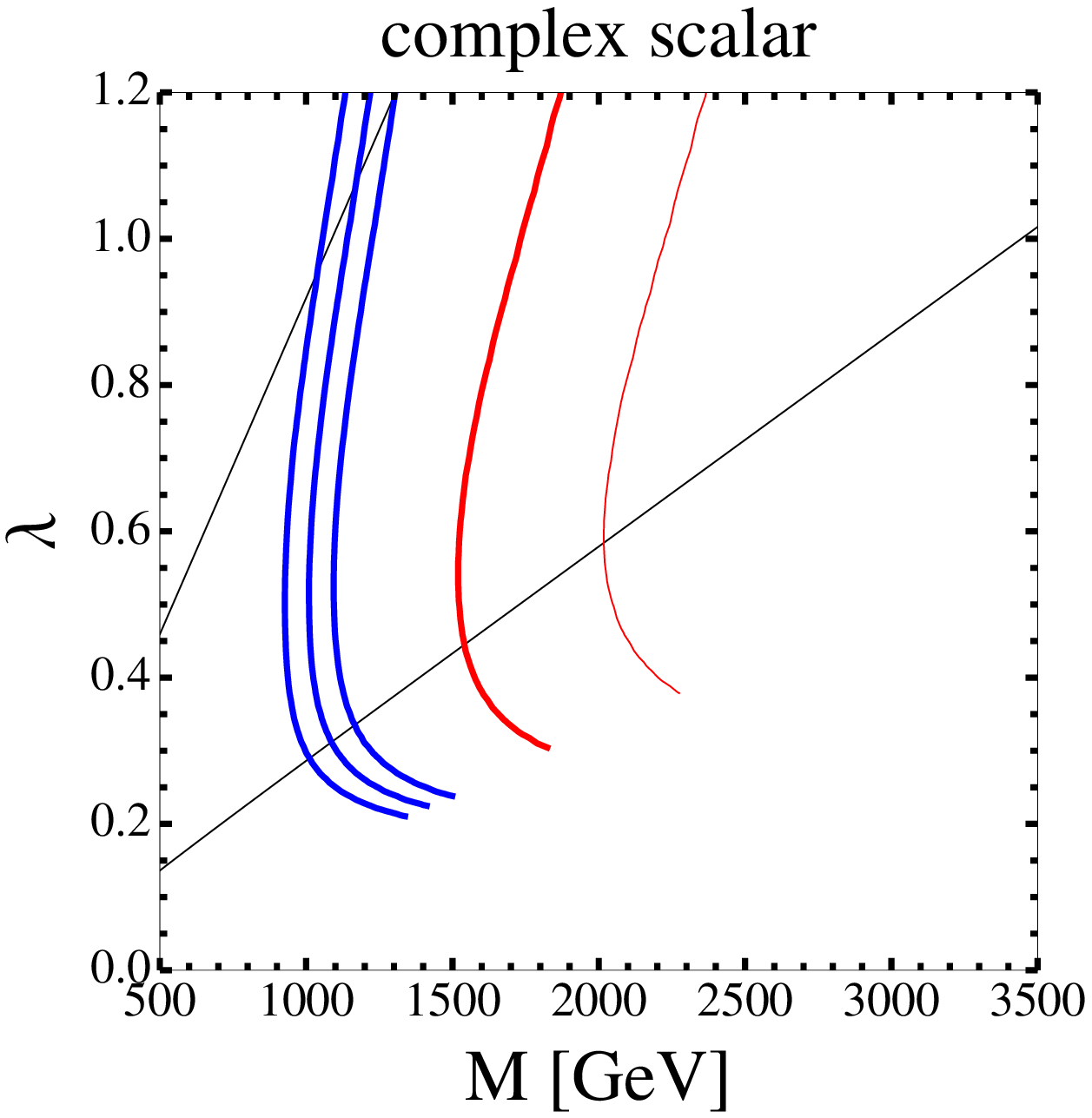}
 \end{center}
 \caption{Contour of the total cross section $\sigma_{\rm tot}$ for
   the pair production of colored mediators, at leading order in
   $(M,\lambda)$ plane, for real and complex scalar
   DM.   Blue lines show the contours
   of $\sigma_{\rm tot}=0.02$, 0.01, and 0.005 pb at $\sqrt{s}=8$ TeV
   from the left to right.  Red lines show the contours of
   $\sigma_{\rm tot}=0.01$ and 0.001 pb at $\sqrt{s}=14$ TeV from the
   left to right.  Note that Yukawa couplings $y_U$ and $y_D$ are
   adjusted through Eqs.~(\ref{eq:relation}), in order to reproduce
   the direct detection cross section. Two black solid lines in each figure show
   the contours of $y_U=0.1$ and 1 from top to bottom.}
\label{fig:y-M}
\end{figure}

In general, DM can couple to all three generations of SM quarks.  In
addition, there are CP phases of the couplings in Eq.~(\ref{eq:L})
which cannot be removed by field redefinitions.  These
flavor--changing and CP--violating couplings are severely
constrained. We will discuss these issues in detail in
Section~\ref{SecFlavor}.  In this and next subsection, we assume that
the couplings to the first generation are dominant, and neglect the
effects of the couplings to the second and third generations.  (We
will omit the generation index $i$ from the coupling constants to the
first generation quarks until Section~\ref{sec:KK}.) For the study of
the signals at LHC, for simplicity, we take the following
parametrization:
\begin{align}
& M_Q = M_U = M_D \equiv M\,,
\label{EqMassassumption}\\
& \lambda_Q = \lambda_U = \lambda_D \equiv \lambda\,,
\\
& y_{U_L} = y_{U_R}  \equiv y_U\,,
\\
& y_{D_L} = y_{D_R}  \equiv y_D\,,
\label{eq:y}
\end{align}
where all parameters are taken to be real.
Then, the effective coupling constants in Table~\ref{tab:operators}
are given by
\begin{align}
C_u^{(\text{R})} = \frac{2\lambda^2 y_U v}{M^2-y_U^2 v^2},~~
C_d^{(\text{R})} = \frac{2\lambda^2 y_D v}{M^2-y_D^2 v^2},~~
C_u^{(\text{Cs})} = \frac{\lambda^2 y_U v}{M^2-y_U^2 v^2},~~
C_d^{(\text{Cs})} = \frac{\lambda^2 y_D v}{M^2-y_D^2 v^2}.
\label{eq:relation}
\end{align}
In our
analysis, we take $C_u^{(\text{R,Cs})} \simeq (68~\text{TeV})^{-1}$
(see Eq.~(\ref{eq:Cu1})).  In Fig.~\ref{fig:y-M}, on $(M, \lambda)$
plane, we show contours on which we obtain $C_u^{(\text{R,Cs})}=
(68~\text{TeV})^{-1}$, taking  $y_U=0.1$ and $1$.  As one can see, the masses of the vector-like quarks must be
$O(1)~\text{TeV}$ as far as all the coupling constants are within the
perturbative regime.

\subsection{Direct production of colored mediators at LHC}

Now we are at the position to discuss the LHC constraints/prospects of
the colored mediators.  In the present scenario, colored mediators are
pair-produced at LHC.  Here, there is an important difference between
the real and complex scalar DM scenarios.  In the former case,
the processes $pp\rightarrow {\cal Q} \bar{\cal Q}$ and ${\cal Q} {\cal
  Q}$ both occur, where ${\cal Q}$ collectively denotes colored
mediators while $\bar{\cal Q}$ is the anti-particles.  
In the case of complex scalar DM, on the contrary,
$pp\rightarrow {\cal Q} {\cal Q}$ is forbidden, so that the relevant
processes are only the pair production of ${\cal Q}$ and $\bar{\cal
  Q}$.
Notice that the amplitudes with
$t$-channel exchange of DM can enhance the cross section in the present scenario.

Once produced, the colored mediators decay into the SM quarks and the
DM particle, so the important processes are~\footnotemark
\footnotetext{
  The process $pp\to {\cal Q}\phi +j$ also contributes to the events with two jets plus missing energy.
  Transverse momenta of the emitted jets tend to be smaller than that given by the pair productions in this process.
  Therefore, the contribution becomes sub-dominant with tighter $p_T$ cuts used in Ref.~\cite{atlas-sq}.
  Since we have neglected these processes, the above bounds are conservative.
}
\begin{align}
  pp\to 
  \left\{
    \begin{array}{l}
      {\cal Q} \bar{\cal Q} \to q \phi^{(*)}\; \bar{q} \phi,
      \\
      {\cal Q} {\cal Q} \to q \phi\; q\phi ~~~
      \mbox{(only for real scalar DM)},
      \\
      \bar{\cal Q} \bar{\cal Q} \to \bar{q} \phi\; \bar{q}\phi ~~~
      \mbox{(only for real scalar DM)},
    \end{array}
  \right.
\end{align}
where $q$ denotes SM quarks while $\bar{q}$ is its anti-particle.
Thus, the LHC signal is two jets plus missing energy.  In
Fig.~\ref{fig:y-M}, we show the contour of total cross section
$\sigma_{\rm tot}$ for the pair production of the colored mediators
for $\sqrt{s}=8\ {\rm TeV}$,
where $\sigma_{\rm tot}$ is calculated by {\sc MadGraph 5}~\cite{Alwall:2011uj} at the leading-order and is given by
\begin{align}
  \sigma_{\rm tot} = 
  \left\{
    \begin{array}{ll}
      \sigma(pp\to {\cal Q} \bar{\cal Q})
      + \sigma(pp\to {\cal Q} {\cal Q})
      + \sigma(pp\to \bar{\cal Q} \bar{\cal Q})
      &:~ \mbox{real scalar DM},
      \\
      \sigma(pp\to {\cal Q} \bar{\cal Q})
      &:~ \mbox{complex scalar DM}.
    \end{array}
  \right.
\end{align}

The di-jet signal with missing energy is studied both at ATLAS and
CMS, particularly in the context of supersymmetric (SUSY) models.  In
the ATLAS analysis~\cite{atlas-sq}, a simplified SUSY model is
studied, where only first two generation squarks and the lightest
neutralino are potentially accessible to LHC while all other SUSY
particles (including the gluino) are heavy.  In such a model, the
lower bound on the common squark mass is 780 GeV, corresponding to the
leading-order squark production cross section of 0.013 pb.  In
general, this value cannot be directly compared with the prediction of the present model
because the signal efficiency (i.e., the fraction of signal events
which pass the cuts in Ref.~\cite{atlas-sq}) may be different.  
By using the parton-level analysis with {\sc MadGraph 5}~\cite{Alwall:2011uj}, we
estimated the efficiency for our model as well as that for the simplified SUSY
model with a squark mass of 780 GeV.
Then, we found
that the former is comparable to or larger than the latter.
Thus, we translate the ATLAS constraint on the simplified SUSY
model to derive a conservative bound on the present model; assuming that
$\sigma_{\rm tot}$ should be smaller than $\sim 0.01\ {\rm pb}$
and O(1) couplings $\lambda, y\lesssim 1$,
$M$ is bounded from below as $M\gtrsim 1-1.5$~TeV ($1-1.1$ TeV)
for real (complex) scalar DM,
depending on the coupling $\lambda$.

Before closing this section, let us discuss the future prospects of the present model.  In
Fig.~\ref{fig:y-M}, we also show the contour of $\sigma_{\rm tot}$ at
$\sqrt{s}=14$~TeV.  At 14 TeV LHC the sensitivity of the search with
two-jets plus missing energy may reach $O$(0.003)~pb and $O$(0.001)~pb
or larger, for the integrated luminosities of 300~fb$^{-1}$ and
3000~fb$^{-1}$, respectively~\cite{Cohen:2013xda}. One can see that a
large region of the parameter space, possibly above $M\simeq 3$ TeV (2 TeV)
for the case of real (complex) scalar DM, may be covered at 14 TeV LHC.


\section{Flavor and CP Constraints}
\label{SecFlavor}

In the previous section, we have discussed the LHC phenomenology of the
isospin--violating DM model with colored mediators.  In the
present scenario, the interaction of DM may 
significantly affect flavor and CP observables, 
which give
very stringent constraints on the present model. 
We concentrate on the case with scalar DM, since 
isospin--violating
fermionic DM with colored mediators is stringently constrained by the
the LHC bounds, as shown in Appendix~\ref{s:maj}.

\subsection{Up- and down-quark masses}

First we discuss the radiative correction to the SM Yukawa
coupling constants in the present model.  In particular, we
concentrate on the Yukawa coupling constants of up- and down-quarks
(which we denote $y_u$ and $y_d$) on which the corrections are the
most significant.

If $y_{U_L}$ or $y_{D_L}$ is non-vanishing, there exist
logarithmically-divergent 1-loop contributions to $y_u$ or $y_d$.
Then, the $\beta$-functions of the up-- and down--quark Yukawa
coupling constants become
\begin{align}
  \frac{d y_u}{d \log \mu} & = \frac{1}{8\pi^2}
  \lambda_Q \lambda_U y_{U_L} + \cdots,
  \\
  \frac{d y_d}{d \log \mu} & = \frac{1}{8\pi^2}
  \lambda_Q \lambda_D y_{D_L} + \cdots,
\end{align}
where $\mu$ is the renormalization scale and ``$\cdots$'' are terms
proportional to $y_u$ or $y_d$.  The important point is that the
$\beta$-functions contain terms which are not proportional to the
SM Yukawa coupling constants.  Consequently, the smallness of
$y_u$ and $y_d$ may be affected in particular when the coupling
constants in the DM sector are relatively large. 
As shown in the previous section, the present scenario
requires large values of $\lambda_{Q,U,D}$ and $y_{U,D}$
(cf. Fig.~\ref{fig:y-M}).
Thus, we expect
significant contribution to the up- and down-quark Yukawa coupling
constants from the DM sector.

The low-energy values of the Yukawa coupling constants, which are
directly related to the up- and down-quark masses, are given by
\begin{align}
  y_u (\mu\ll M_Q) & \sim y_u (M_*) 
  + \frac{1}{8\pi^2} \lambda_Q \lambda_U y_{U_L}
  \log \frac{M_Q}{M_*} + \cdots,
  \label{y_u(M_Q)}
  \\
  y_d (\mu\ll M_Q) & \sim y_d (M_*) 
  + \frac{1}{8\pi^2} \lambda_Q \lambda_D y_{D_L}
  \log \frac{M_Q}{M_*} + \cdots,
  \label{y_d(M_Q)}
\end{align}
where $M_*$ is the cut-off scale at which the boundary conditions are
given.  If $\lambda_{Q,U,D}\sim y_{U_L,D_L}\sim 1$, the second terms in Eqs.\
\eqref{y_u(M_Q)} and \eqref{y_d(M_Q)} are estimated to be larger than
$O(10^{-2})$, which is much larger than the SM values of
those Yukawa coupling constants.  In order to realize the Yukawa
coupling constants compatible with the up-- and down--quark masses, such
contributions should be cancelled by $y_{u,d} (M_*)$, which requires
a significant tuning between those two unrelated quantities.

For the scenario of isospin--violating DM, in fact, $y_{U_L}$
and $y_{D_L}$ may vanish; in order to generate the operator $\phi\phi\bar{q}q$, we only need $y_{U_R}$ and $y_{D_R}$.  They also
affect the up- and down-quark Yukawa coupling constants.  The
contributions which are proportional to $y_{U_R}$ and $y_{D_R}$ are
finite, and are given by
\begin{align}
  \Delta y_u & = \frac{1}{8 \pi^2} 
  \lambda_Q \lambda_U y_{U_R}
  \frac{M_Q M_U}{M_Q^2 - M_U^2} \log \frac{M_U}{M_Q},
  \\
  \Delta y_d & = \frac{1}{8 \pi^2} 
  \lambda_Q \lambda_U y_{D_R}
  \frac{M_Q M_D}{M_Q^2 - M_D^2} \log \frac{M_D}{M_Q},
\end{align}
which are still much larger than the SM values of up- and
down-quark Yukawa coupling constant if $\lambda_{Q,U,D} \sim
y_{U_R,D_R}\sim 1$.  Thus, the serious tunings of the counter terms of
the Yukawa coupling constants are unavoidable in the present model.

\subsection{Electric dipole moment of neutron}

Next, we consider the electric dipole moment (EDM) of the neutron.
If the newly introduced coupling constants have phases, which is the
case in general, they become a new source of CP violations.  In the
present model, the DM sector necessarily couple to the first
generation quarks, so the important check point is the neutron EDM.

In order to see how large the neutron EDM becomes, we calculate the
coefficients of the EDM and chromo-EDM (CEDM) operators of up- and
down-quarks:
\begin{align}
  {\cal L}_{\rm (C)EDM} = \frac{i}{2}
  \sum_{f=u,d} 
  \left[ 
    d_f F_{\mu\nu} \bar{f} \sigma_{\mu\nu} \gamma_5 f
    + g_3 \tilde{d}_f G_{\mu\nu}^{(a)} T^a_{\alpha\beta}
    \bar{f}_\alpha \sigma_{\mu\nu} \gamma_5 f_\beta
  \right],
\end{align}
where $F_{\mu\nu}$ and $G_{\mu\nu}^{(a)}$ are field-strength tensors
of photon and gluon, respectively, $g_3$ is the strong gauge coupling
constant, and $T^a_{\alpha\beta}$ is the generator for $SU(3)_C$ (with
$\alpha$ and $\beta$ being color indices, while $a$ being index for
the adjoint representation).  With the (C)EDMs of quarks being given,
the neutron EDM is estimated as \cite{Hisano:2012sc}
\begin{align}
  d_n = - 0.12 d_u + 0.47 d_d 
  + e ( -0.18 \tilde{d}_u + 0.18 \tilde{d}_d ).
\end{align}
(The numerical uncertainties in QCD parameters may change the above
formula by $\sim 10\ \%$ \cite{Hisano:2012sc}.  The conclusion of this
subsection is, however, unaffected by such an uncertainty.)

As shown in the previous section, the LHC bounds require that the
masses of the colored mediators should be much larger than the Higgs
VEV, $M_{Q,U,D} > v$.  In such a case, the coefficients of the (C)EDM
operators can be expanded in powers of the Higgs VEV, and we only keep
the leading-order terms in $v$.  In the limit of $m_\phi\ll M_{Q,U,D}$
(with $m_\phi$ being the mass of the scalar DM) we obtain
\begin{align}
  d_u & = \frac{1}{32 \pi^2} \frac{e Q_U v}{M_Q M_U}
  \Im (\lambda_Q \lambda_U^* y_{U_R}),
  \\
  \tilde{d}_u & = \frac{1}{32 \pi^2} \frac{v}{M_Q M_U}
  \Im (\lambda_Q \lambda_U^* y_{U_R}),
\end{align}
and $d_d$ and $\tilde{d}_d$ are obtained from $d_u$ and $\tilde{d}_u$
by replacing the subscripts as $U\rightarrow D$.  Here, $e$ is the
electric charge, $Q_U=\frac{2}{3}$, and $Q_D=-\frac{1}{3}$.  We note
here that, at the leading order in $v$, the contribution proportional
to $\Im (\lambda_Q \lambda_U^* y_{U_L})$ vanishes.

Taking $M_Q=M_U=M_D$ for simplicity, we obtain
\begin{align}
  d_n \simeq \left[
    -2.8 \times 10^{-21} e\ {\rm cm} 
    \times \Im (\lambda_Q \lambda_U^* y_{U_R})
    + 
    2.5 \times 10^{-22} e\ {\rm cm} 
    \times \Im (\lambda_Q \lambda_D^* y_{D_R})
  \right]
  \left( \frac{1\ {\rm TeV}}{M_Q} \right)^2.
\end{align}
This should be compared with the present bound on the neutron EDM, which is given by
\cite{Beringer:1900zz}
\begin{align}
  |d_n| < 0.29 \times 10^{-25}  e\ {\rm cm}.
\end{align}
Thus, the neutron EDM provides a very severe constraint on 
the complex phases of the couplings,
$\Im (\lambda_Q \lambda_{U}^* y_{U_R})\lesssim O(10^{-5}-10^{-4})$
and
$\Im (\lambda_Q \lambda_{D}^* y_{D_R})\lesssim O(10^{-4}-10^{-3})$,
for $M_Q\simeq O(1-3)$ TeV.

\subsection{$K$-$\bar{K}$ mixing}
\label{sec:KK}

In the present analysis, we introduced only one set of vector-like
fermions (i.e., $Q$, $U$, and $D$) for minimality.  No symmetry
forbids their interactions with second- and third-generation quarks.
Such interactions in general induce unwanted CP and flavor violations;
it is often the case that the $K$-$\bar{K}$ mixing parameters, i.e.,
$\epsilon_K$ and $\Delta m_K$, give stringent constraints.  Thus, we
consider them in this subsection.

The effective $\Delta S=2$ Hamiltonian can be described as
\begin{align}
  {\mathcal H}_{\rm eff} = \sum_{i=1}^3 
  \left( C_{L,i} {\mathcal Q}_{L,i}
  + C_{R,i} {\mathcal Q}_{R,i} \right)
  + \sum_{i=4}^5 C_i {\mathcal Q}_i,
  \label{eq:Heff_DelS2}
\end{align}
where the operators are 
\begin{align}
  {\mathcal Q}_{L,1} &= ( \bar{d}_\alpha \gamma_\mu P_L s_\alpha )
  ( \bar{d}_\beta \gamma^\mu P_L s_\beta) ,
  \\
  {\mathcal Q}_{L,2} &= ( \bar{d}_\alpha P_L s_\alpha )
  ( \bar{d}_\beta P_L s_\beta ) ,
  \\
  {\mathcal Q}_{L,3} &= ( \bar{d}_\alpha P_L s_\beta )  
  ( \bar{d}_\beta P_L s_\alpha ),
  \\
  {\mathcal Q}_4 &= ( \bar{d}_\alpha P_L s_\alpha) 
  ( \bar{d}_\beta P_R s_\beta ),
  \\
  {\mathcal Q}_5 &= ( \bar{d}_\alpha P_L s_\beta )  
  ( \bar{d}_\beta P_R s_\alpha ),
  \label{eq:DelS2OP}
\end{align}
and ${\mathcal Q}_{R,i}=[{\mathcal Q}_{L,i}]_{L\rightarrow R}$.  We
calculate the Wilson coefficients in the present model.  As in the
case of neutron EDM, we use the mass-insertion approximation and only
consider the leading contributions with respect to the insertions of
the Higgs VEV.

In the case of real scalar DM, sum of the diagrams with no
Higgs-VEV insertion vanishes, and the leading contributions are given
by
\begin{align}
  C_{L,2}^{(\phi: \rm real)} & =  \frac{1}{16\pi^2}
  \frac{v^2}{M_Q^4} (\lambda_D^{s*} \lambda_Q^{d})^2 
  \left[
    y_{D_R}^2 F_0 (x_D, x_\phi) + 
    y_{D_R} y_{D_L} F_1 (x_D) + 
    y_{D_L}^2 F_2 (x_D)
  \right],
  \\
  C_{R,2}^{(\phi: \rm real)} & =  \frac{1}{16\pi^2}
  \frac{v^2}{M_Q^4} (\lambda_Q^{s*} \lambda_D^{d})^2 
  \left[
    y_{D_R}^{*2} F_0 (x_D, x_\phi) + 
    y_{D_R}^* y_{D_L}^* F_1 (x_D) + 
    y_{D_L}^{*2} F_2 (x_D)
  \right],
  \\
  C_4^{(\phi: \rm real)} & =  \frac{1}{8\pi^2}
  \frac{v^2}{M_Q^4} 
  \lambda_Q^{s*} \lambda_D^{s*} 
  \lambda_Q^{d}  \lambda_D^{d}
  \left[
    y_{D_R} y_{D_R}^* F_0 (x_D, x_\phi) + 
    \Re (y_{D_R} y_{D_L}^*) F_1 (x_D) + 
    y_{D_L} y_{D_L}^* F_2 (x_D)
  \right],  
\end{align}
where
\begin{align}
  F_0 (x_D, x_\phi) &=
  - \frac{\log x_\phi}{x_D}
  + \frac{-2 x_D^3 + 4 x_D^2 - 4 x_D + (3 x_D-1) \log x_D + 2}
  {(x_D-1)^3 x_D},
  \\
  F_1 (x_D) &=
  \frac{-2x_D^2 + 4 x_D \log x_D + 2}{\sqrt{x_D} (x_D-1)^3},
  \\
  F_2 (x_D) &=
  \frac{-2 x_D + (x_D+1) \log x_D+2}{(x_D-1)^3},
\end{align}
with $x_D\equiv M_D^2/M_Q^2$ and $x_\phi\equiv m_\phi^2/M_Q^2$.  (The
superscripts $d$ and $s$ of $\lambda_{Q,D}^{d,s}$ denote the coupling
constants to the first and second generations, respectively,
cf. Eq.~(\ref{eq:L}).)  Notice that the above expressions are valid
only when $m_\phi\ll M_{Q,D}$.  (Other Wilson coefficients vanish at
this order.)  For complex scalar DM, we obtain
\begin{align}
  C_{L,1}^{(\phi: \rm complex)} = & - \frac{1}{128 \pi^2} \frac{1}{M_Q^2}
  (\lambda_Q^{s*} \lambda_Q^{d})^2,
  \\
  C_{R,1}^{(\phi: \rm complex)} = & - \frac{1}{128 \pi^2} \frac{1}{M_D^2}
  (\lambda_D^{s*} \lambda_D^{d})^2,
  \\
  C_{L,2}^{(\phi: \rm complex)} = &  \frac{1}{32\pi^2}
  \frac{v^2}{M_Q^4} (\lambda_D^{s*} \lambda_Q^{d})^2 
  \left[
    y_{D_R}^2 F_0 (x_D, x_\phi) + 
    y_{D_R} y_{D_L} F_1 (x_D) + 
    y_{D_L}^2 F_2 (x_D)
  \right],
  \\
  C_{R,2}^{(\phi: \rm complex)} = &  \frac{1}{32\pi^2}
  \frac{v^2}{M_Q^4} (\lambda_Q^{s*} \lambda_D^{d})^2 
  \left[
    y_{D_R}^{*2} F_0 (x_D, x_\phi) + 
    y_{D_R}^* y_{D_L}^* F_1 (x_D) + 
    y_{D_L}^{*2} F_2 (x_D)
  \right],
  \\
  C_4^{(\phi: \rm complex)} = &  \frac{1}{16\pi^2}
  \frac{v^2}{M_Q^4} 
  \lambda_Q^{s*} \lambda_D^{s*} 
  \lambda_Q^{d}  \lambda_D^{d}
  \nonumber \\ &
  \left[
    y_{D_R} y_{D_R}^* F_0 (x_D, x_\phi) + 
    \Re (y_{D_R} y_{D_L}^*) F_1 (x_D) + 
    y_{D_L} y_{D_L}^* F_2 (x_D)
  \right],
  \\
  C_5^{(\phi: \rm complex)} = & \frac{1}{16 \pi^2}
  \lambda_Q^{s*} \lambda_D^{s*} \lambda_Q^{d} \lambda_D^{d}
  \frac{1}{M_Q^2 - M_D^2} \log \frac{M_Q}{M_D},
\end{align}
where we neglected the terms which are higher order in $v$.  (Other
Wilson coefficients vanish at this order.)

With the Wilson coefficients, we calculate the matrix elements
relevant for the study of $K$-$\bar{K}$ mixing parameters.  Here, our
purpose is to obtain semi--quantitative bounds on the model parameters, so
we use the vacuum-insertion approximation to evaluate the matrix
elements.  Then, we obtain \cite{Buras:2001ra}
\begin{align}
  \langle K | {\cal H}_{\rm eff} | \bar{K} \rangle = &
  \frac{2}{3} (m_K f_K)^2 \left( C_{L,1} + C_{R,1} \right) 
  - \frac{5}{12} \frac{m_K^2}{m_s^2}
  (m_K f_K)^2 \left( C_{L,2} + C_{R,2} \right)
  \nonumber \\ &
  + \frac{1}{12} \frac{m_K^2}{m_s^2}
  (m_K f_K)^2 \left( C_{L,3} + C_{R,3} \right)
  + \left(
    \frac{1}{12} + \frac{1}{2}
    \frac{m_K^2}{m_s^2}
  \right)
  (m_K f_K)^2 C_{4}
  \nonumber \\ &
  + \left(
    \frac{1}{4} + 
    \frac{1}{6} \frac{m_K^2}{m_s^2}
  \right)
  (m_K f_K)^2 C_{5},
\end{align}
where $m_K$ is the mass of $K$, $m_s\simeq 95\ {\rm MeV}$ is the
strange-quark mass, and $f_K\simeq 160\ {\rm MeV}$ is the decay
constant.  With the above matrix element, we estimate the
DM sector contributions to the $K$-$\bar{K}$ mixing
parameters as
\begin{align}
  |\epsilon_K^{(\phi)}| &= 
  \frac{\Im \langle K | {\cal H}_{\rm eff} | \bar{K} \rangle}
  {2\sqrt{2} m_K \Delta m_K },
  \label{eps_K}
  \\
  \Delta m_K^{(\phi)} &= 
  \frac{1}{m_K}
  |\langle K | {\cal H}_{\rm eff} | \bar{K} \rangle|.
\end{align}
The numerical values of $\epsilon_K^{(\phi)}$ and $\Delta
m_K^{(\phi)}$ depend on various parameters.  Taking
\begin{align}
  & \lambda_Q^s = \lambda_D^2 \equiv \lambda^s,
  \\
  & y_{D_L} = y_{D_R} \equiv y_{D},
\end{align}
as well as the relations given in Eqs.\ \eqref{EqMassassumption} --
\eqref{eq:y}, for example, we obtain
\begin{align}
  |\epsilon_K^{(\phi: \rm real)}| \simeq 
  6.5\times 10^3
  \times \Im (\lambda^{s*} \lambda^d)^2 y_D^2
  \lrfp{1~{\rm TeV}}{M}{4}
  \left[ 1 + 0.34 \log \frac{(M/m_\phi)}{100} \right],
  \label{epsk(real)}
\end{align}
and
\begin{align}
  |\epsilon_K^{(\phi: \rm complex)}| \simeq &
  1.7\times 10^4
  \times \Im (\lambda^{s*} \lambda^d)^2 
   \lrfp{1~{\rm TeV}}{M}{2}
  \nonumber \\ &
  + 3.2\times 10^3
  \times \Im (\lambda^{s*} \lambda^d)^2 y_D^2
   \lrfp{1~{\rm TeV}}{M}{4}
  \left[ 1 + 0.34 \log \frac{(M/m_\phi)}{100} \right],
  \label{epsk(complex)}
\end{align}
for the cases where $\phi$ is real and complex, respectively.
  (Here, we assumed that $y_D$ is
real for simplicity.)  In addition, with the present choice of
parameters,
\begin{align}
  \Delta m_K^{(\phi)} \simeq 
  1.5\times 10^{10}\ {\rm sec}^{-1}
  \times 
  \frac{|\lambda^{s*} \lambda^d|^2}{\Im (\lambda^{s*} \lambda^d)^2}
  |\epsilon_K^{(\phi)}|.
  \label{dmk}
\end{align}

The measured values of the $K$-$\bar{K}$ mixing parameters are well
explained by the SM, and there exist stringent constraints
on the extra contributions to those quantities.  Comparing the
SM prediction ($\epsilon_K^{\rm (SM)} = (1.81\pm 0.28)
\times 10^{-3}$ \cite{Brod:2011ty}) and the experimental value
($\epsilon_K^{\rm (exp)} = (2.228\pm 0.011) \times 10^{-3}$
\cite{Beringer:1900zz}), the DM sector contribution to
$\epsilon_K$ is constrained to be $|\epsilon_K^{(\phi)}| <9.8\times
10^{-4}$.  In addition, the experimental value of $\Delta m_K$ is
known to be $\Delta m_K^{\rm (exp)}=(0.5293\pm 0.0009)\times 10^{10}\
{\rm sec}^{-1}$ \cite{Beringer:1900zz}, which we use as an upper bound on
$\Delta m_K^{(\phi)}$.  Assuming no accidental cancellation among
contributions from different Feynman diagrams, the DM sector
contributions are likely to become much larger than the upper bounds
on those quantities unless some of the coupling constants are much
smaller than $1$, as indicated by Eqs.\ \eqref{epsk(real)} --
\eqref{dmk}.

\section{Conclusion}
\label{sec:conclusion}

In this paper, we have studied isospin--violating light DM
that can explain the possible CDMS-Si signal of light DM,
while avoiding the constraints by recent LUX and SuperCDMS experiments.
In particular, we considered isospin--violating light DM models with colored mediators.
We have shown that a minimal viable model includes
 scalar DM and new colored vector-like fermions with
masses of $O(1)$ TeV as mediators.  We investigated the collider
searches, flavor and CP phenomenology.
The masses of colored mediators are constrained by the 8 TeV LHC results
as $M\gtrsim 1-1.5$ TeV ($1-1.1$ TeV) for real (complex) scalar DM. 
The 14 TeV LHC may cover a large region of the remaining parameter space.

We have also studied flavor and CP constraints on the colored-mediator
model for the isospin--violating DM.  In such a model, the
interaction of quarks with colored mediator and DM should be
sizable, which results in large radiative correction to flavor and CP
observables.  We have studied the effects on the quark masses (in
particular, those of up- and down-quarks), EDM of neutron, and the
$K$-$\bar{K}$ mixing parameters.  
Radiative corrections to the SM Yukawa couplings from the DM sector
are extremely large, and hence fine--tunings are unavoidable.
Flavor and CP violating observables also impose severe constraints 
on the present scenario.

\section*{Acknowledgment}

This work was supported by JSPS KAKENHI Grant 
No.~22244021 (K.H., T.M.), No.~22540263 (T.M.), No.~23104008 (T.M.)
and also by World Premier International Research Center Initiative
(WPI Initiative), MEXT, Japan. The work of S.P.L. was supported by the
Program for Leading Graduate Schools, MEXT, Japan.
The work of Y.Y. has been supported in part by the Ministry of Economy and Competitiveness (MINECO), grant FPA2010-17915, and by the Junta de Andaluc\'ia, grants FQM 101 and FQM 6552.

\appendix
\section{Isospin--Violating Fermionic Dark Matter with Colored Scalar Mediators}
\label{s:maj}
In this appendix, we briefly discuss isospin--violating fermionic DM models with colored scalar mediators.
The effective operators $\mathcal{O}^\text{(M)}$ and $\mathcal{O}^\text{(Ds)}$ in Table~\ref{tab:operators} can be induced by exchanges of colored scalars
$\tilde{Q}_L$, $\tilde{Q}_R$ with the following Lagrangian:
\begin{equation}
  L \supset y_L\chi q_L\tilde{Q}_L +  y_R\chi q_R\tilde{Q}_R
  + A H \tilde{Q}_L^*\tilde{Q}_R.
\end{equation}
The benchmark point in Eq.~(\ref{eq:Cu2}),
when interpreted with this Lagrangian,
corresponds to
\begin{align}
  C^{\text{(M)}}
  \simeq &
  \frac{y_L y_R A v}{2 m_{\tilde{Q}}^4}
  \simeq 
  \frac{1}{(1.05\ \rm{TeV})^2} ,
  \\
  C^{\text{(Ds)}}
  \simeq &
  \frac{y_L y_R A v}{4 m_{\tilde{Q}}^4}
  \simeq 
  \frac{1}{(1.05\ \rm{TeV})^2} ,
\end{align}
where we assume all colored scalars have common mass $m_{\tilde{Q}}$.
Assuming
that $y_L$, $y_R\lesssim 1$, and $A \lesssim m_{\tilde{Q}}$ for
perturbative unitarity condition, the colored scalar mass parameter
should be smaller than 460 GeV (360 GeV) for Majorana (Dirac)
DM. 
 
If the colored scalar is produced at the LHC, it will decay into a SM
quark and DM.  This collider signature is analogous to that
of SUSY models with almost massless neutralino and a very heavy
gluino.  Such a simplified SUSY model is searched for at the LHC, and
the lower limit on the mass of squark is 780 GeV~\cite{atlas-sq}.  The
limit can be directly applied to the current setup, since squark pairs
are mainly produced by QCD processes in both models. 
Hence, as an explanation of the CDMS-Si signal,
isospin--violating fermionic DM models with colored scalar mediators
are already disfavored by current LHC results.

\end{document}